\documentclass{sf2a-conf2024}
\usepackage{graphicx}
\usepackage{hyperref}
\usepackage[]{natbib}  
\usepackage{epstopdf}

\def\BibTeX{{\rm B\kern-.05em{\sc i\kern-.025em b}\kern-.08em
    T\kern-.1667em\lower.7ex\hbox{E}\kern-.125emX}}
\bibpunct{(}{)}{;}{a}{}{,}  

\begin{document}

\TitreGlobal{SF2A 2024}


\title{"My Earth'' Astrophysics and Planets - a serious game to build low-carbon scenarios in the astronomy academic community}

\runningtitle{"My Earth'' Astrophysics and Planets''}

\author{F.~Malbet}
\address{Univ.~Grenoble Alpes, CNRS, IPAG, Grenoble, France}
\author{A. Santerne}
\address{Aix Marseille Univ., CNRS, CNES, LAM, Marseille, France}
\author{J.~ Milli$^1$} 
\author{N.~Champollion}\address{Univ.~Grenoble Alpes, CNRS, Grenoble-INP,
  INRAE, IRD, IGE, Grenoble, France}
\author{L.~Lamy$^2$} 
\author{H.~Imbaud}\address{Aix Marseille Univ., CNRS, CINAM,
  Marseille, France}
\author{F.~Gaunet}\address{Aix Marseille Univ., CNRS, CRPN,
  Marseille, France}
\author{T.~Masson}\address{
  Aix Marseille Univ, Université de Toulon, CNRS, CPT, Marseille, France}
\author{A.-M.~Daré}\address{Aix Marseille Univ., CNRS, Univ. Toulon, IM2NP,
  Marseille, France}
\author{N.~Gratiot} 
\author{P.~Bellemain}\address{Univ.~Grenoble Alpes, CNRS,
  Grenoble-INP, GIPSA-lab, Grenoble, France}
\setcounter{page}{237}


\maketitle


\begin{abstract}
  This report summarizes what has happened in the mini-workshops
  entitled "\emph{My Earth in 180 minutes}'' organized during the
  lunch break at the SF2A 2024 conference in Marseille. The
  project showcased an innovative serious game designed to raise
  awareness of greenhouse gas (GHG) emissions in astronomical research
  laboratories. Participants, organized into teams, simulate
  strategies to reduce their carbon footprints by 50\%, focusing on
  key astronomical activities such as space instrumentation, data
  analysis, and laboratory work. The sessions highlight the challenges
  of achieving significant emissions reductions without disrupting
  core research activities, such as telescope observations. While the
  serious game facilitates important discussions on sustainable practices, the
  results point to the need for broader engagement, adaptation to
  different cultural contexts, and institutional support. The project
  highlights the importance of integrating climate action into the
  academic environment and suggests potential future directions for
  expanding its impact.
\end{abstract}



\section{Introduction}

During the French Week of Astrophysics (SF2A) held in Marseille from
June 4 to June 7, 2024, seven mini-workshops were organized during the
lunch break following the methodology \textit{My Earth in 180 Minutes}
created and developed by several researchers in Grenoble
\citep{gratiot:hal-04126329, gratiot:theconversation2024}. There are
several versions in various scientific domains. The serious game version used
at this conference is the one that was adapted by \citet{malbet:hal-04732729}
for \emph{Astrophysics and Planets} laboratory teams.

This project emerged in the context of the global climate crisis and
the urgent need to reduce greenhouse gas (GHG) emissions in research
laboratories, particularly in the field of
astronomy\citep{2022NatAs...6..503K, 2024ccac.book...18K}. The
\emph{My Earth in 180 Minutes} serious game is part of a series of
initiatives aimed at raising awareness in the scientific community
about the urgency of changing professional practices. This report
summarises the objectives, the process and the results of this
initiative, which were presented and experienced during the SF2A~2024
conference.

\section{Objectives and motivation}

The main objectives of this game-based transition support system are
(1) to raise awareness among astronomy researchers about the
environmental impact of their professional practices; (2) to stimulate
the development of emission reduction trajectories within research
labs; and (3) to test innovative strategies through a serious game,
enabling participants to simulate the effects of specific actions on
their CO2 emissions.

The project initiators highlight that monitoring emissions, while
essential, is not sufficient to deeply change behavior. By immersing
participants with their pairs in simulations where they must adopt roles and strategies
different from their usual practices, the serious game aims at provoking deeper
reflection and commitment.

\section{The serious game “My Earth in 180 Minutes”}

\begin{figure}[t]
  \centering
  \fbox{\includegraphics[width=0.55\textwidth]{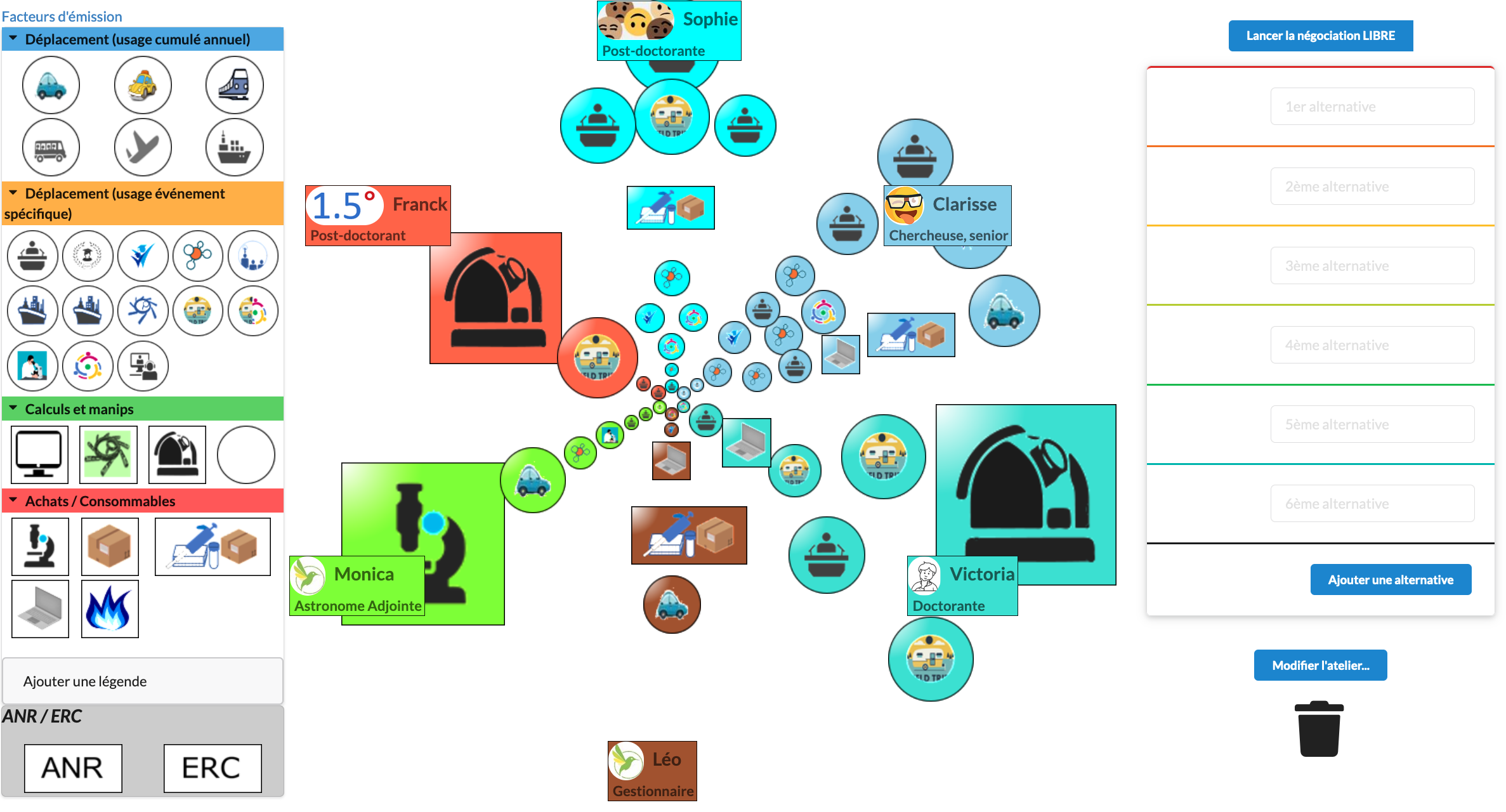}}
  \fbox{\includegraphics[width=0.4\textwidth]{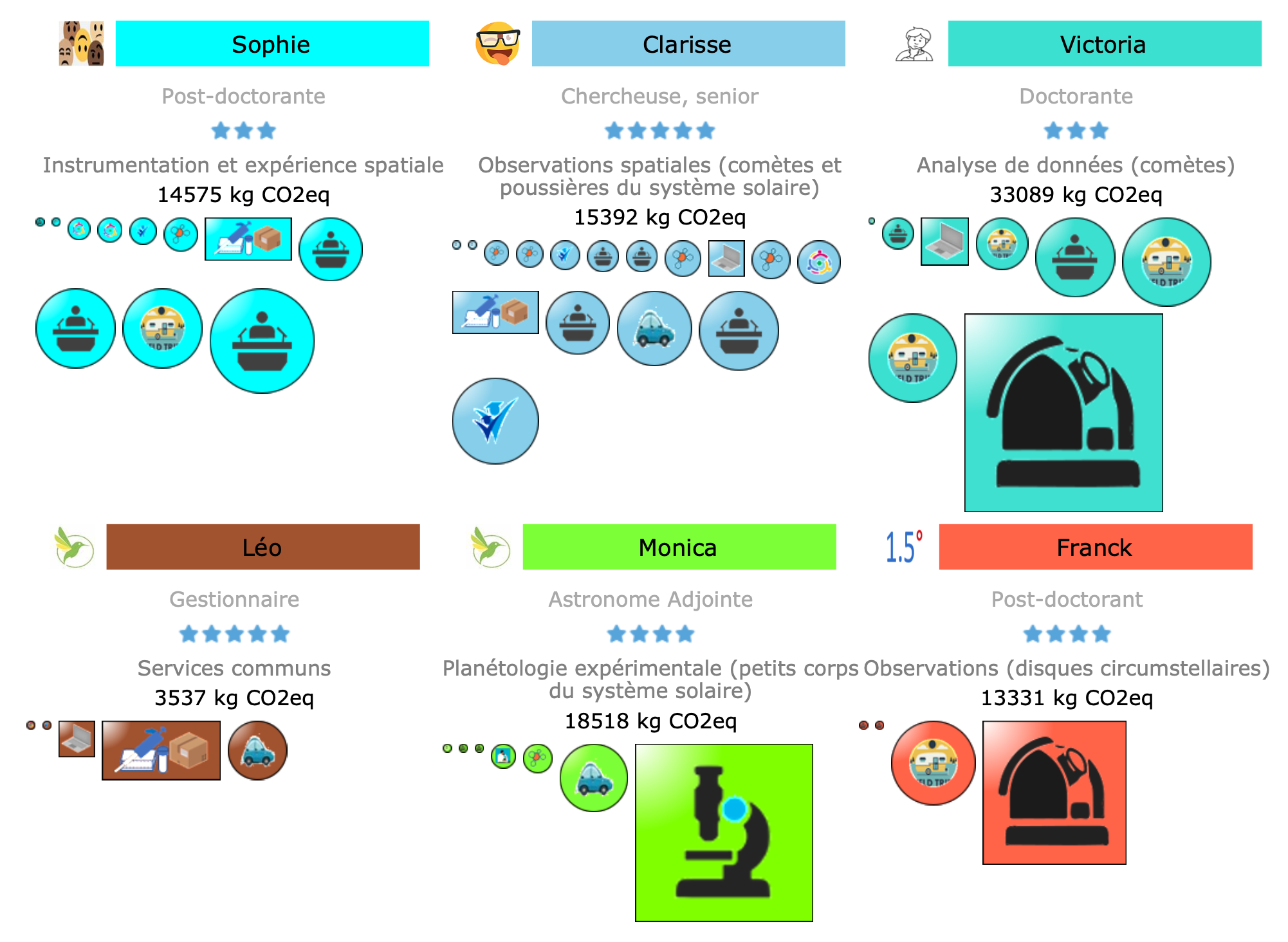}}
  \caption{Left panel : "\emph{Astrophysics and planets}” game board. Right
    panel : "\emph{Astrophysics and planets}” characters.}
  \label{fig:board}
\end{figure}

The serious game provides a simulation framework in which
participants, working in scientific teams, must develop strategies to
reduce their GHG emissions by negociating while maintaining the
efficiency and productivity of their research. Participants are
invited to project themselves into the skin of a fictional character
with specific statuts, gender, carreer, psychological profile.  The
scenario is built around three main astronomical activities : space
instrumentation, laboratory work, data observation and simulation. One
should remember that the average carbon emission per astronomer has
recently been estimated at about 37 tons CO2eq
\citep{2022NatAs...6..503K}.

A concrete example of the simulation was shared during the session in
Marseille. Each participant or team embodies a specific role, such as
a PhD student, a postdoctoral researcher, an astronomer, or an
administrator, with associated initial emissions
(Fig.~\ref{fig:board}). For instance, a third-year PhD student is
responsible for 33.1 tons CO2eq, while a senior scientist generates
around 15.4 tons CO2eq.

The sessions of this game-based transition support system include
negotiation phases where participants aim at achieving a 50\% emissions
reduction. Various actions are possible: reducing professional
travels, extending equipment lifespan, prioritizing the use of
archived data over new observations, or pooling long-distance
travels. These actions are not specific, and playing a role allows
creativity, imagination, and even encourages it, so as to widen the
field of possibilities. 

\section{Results from the SF2A mini-workshops}

\begin{figure}[t]
  \centering 
  \includegraphics[width=0.3\textwidth]{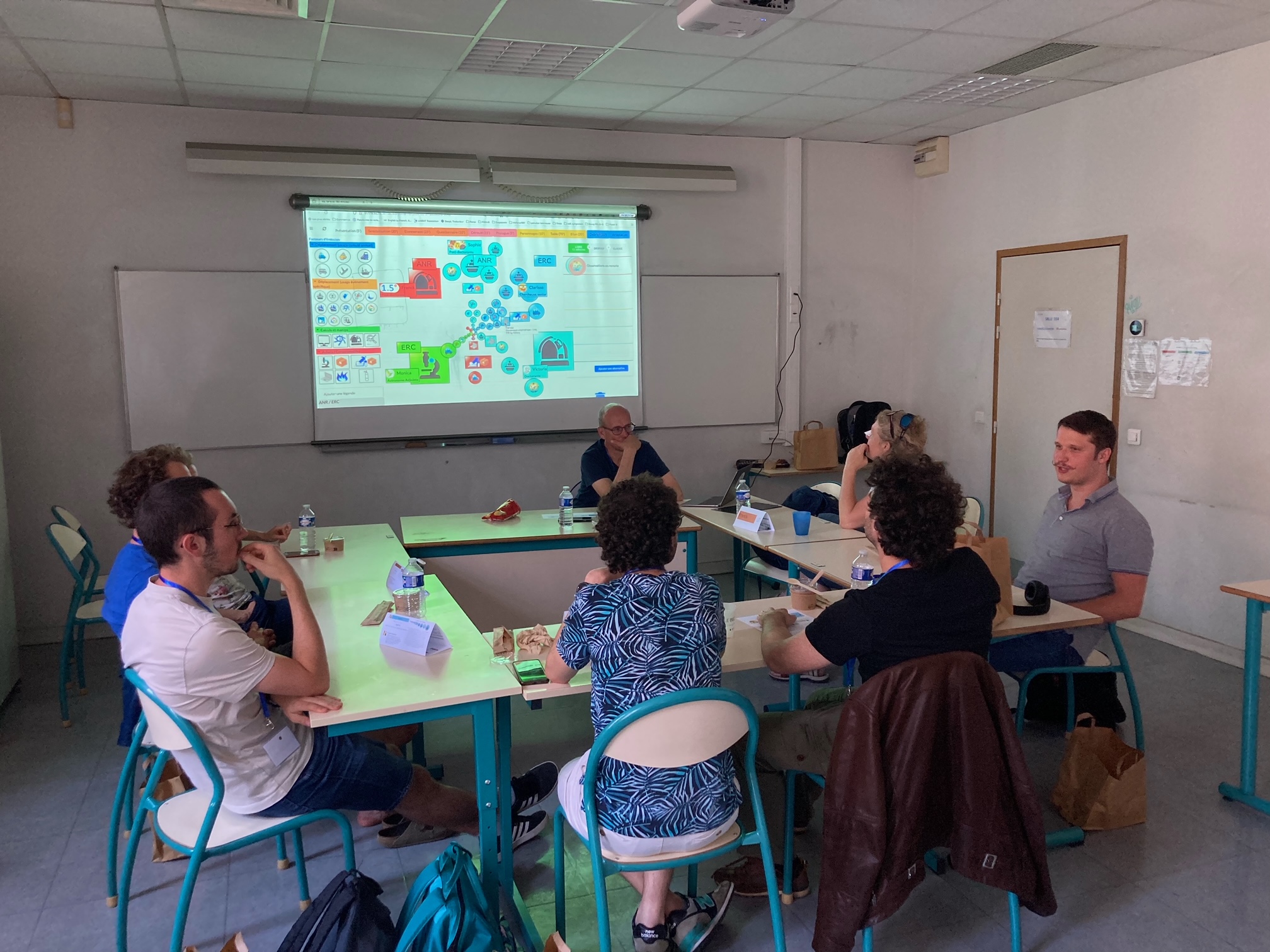}
  \includegraphics[width=0.3\textwidth]{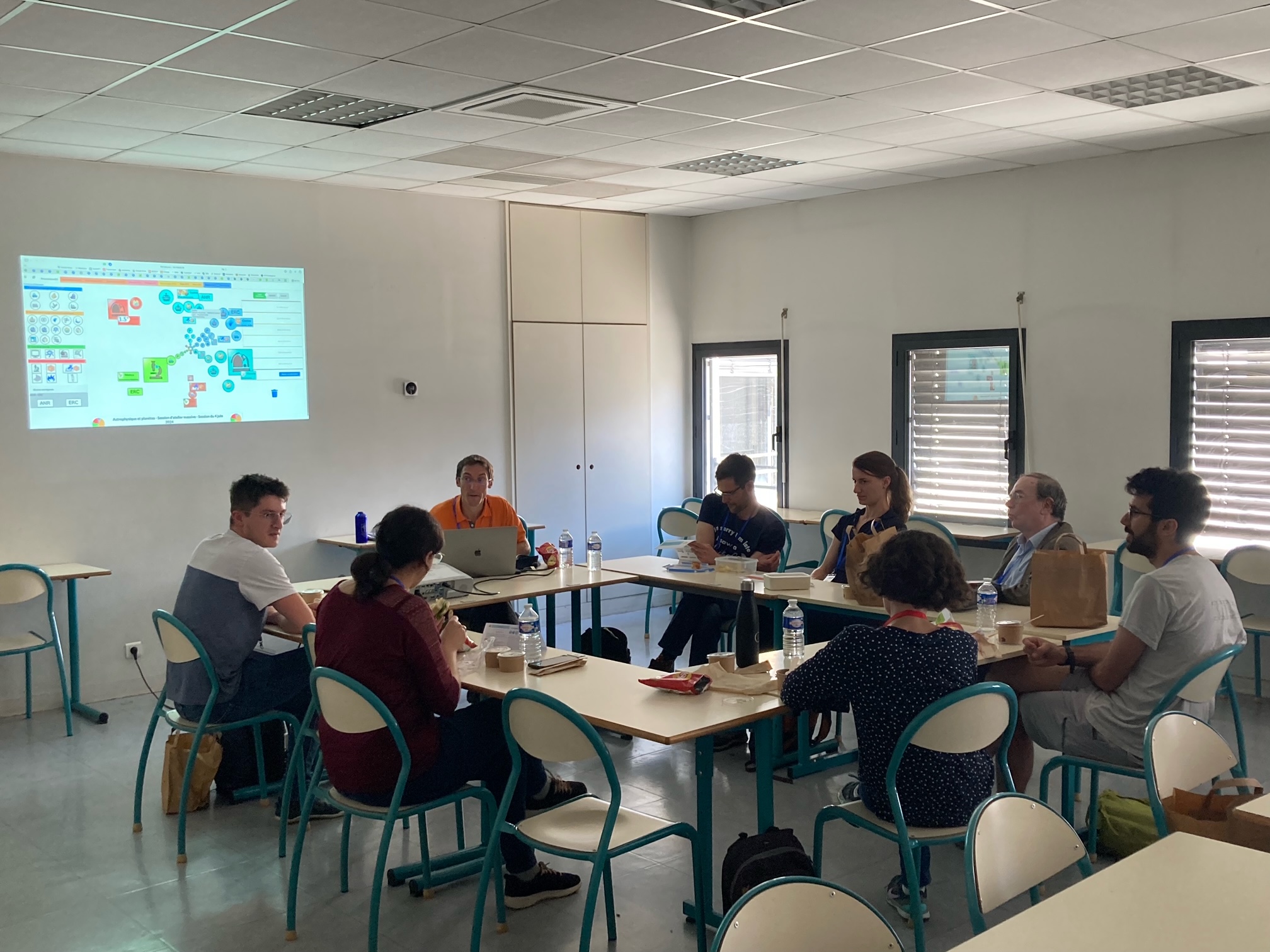}
  \includegraphics[width=0.3\textwidth]{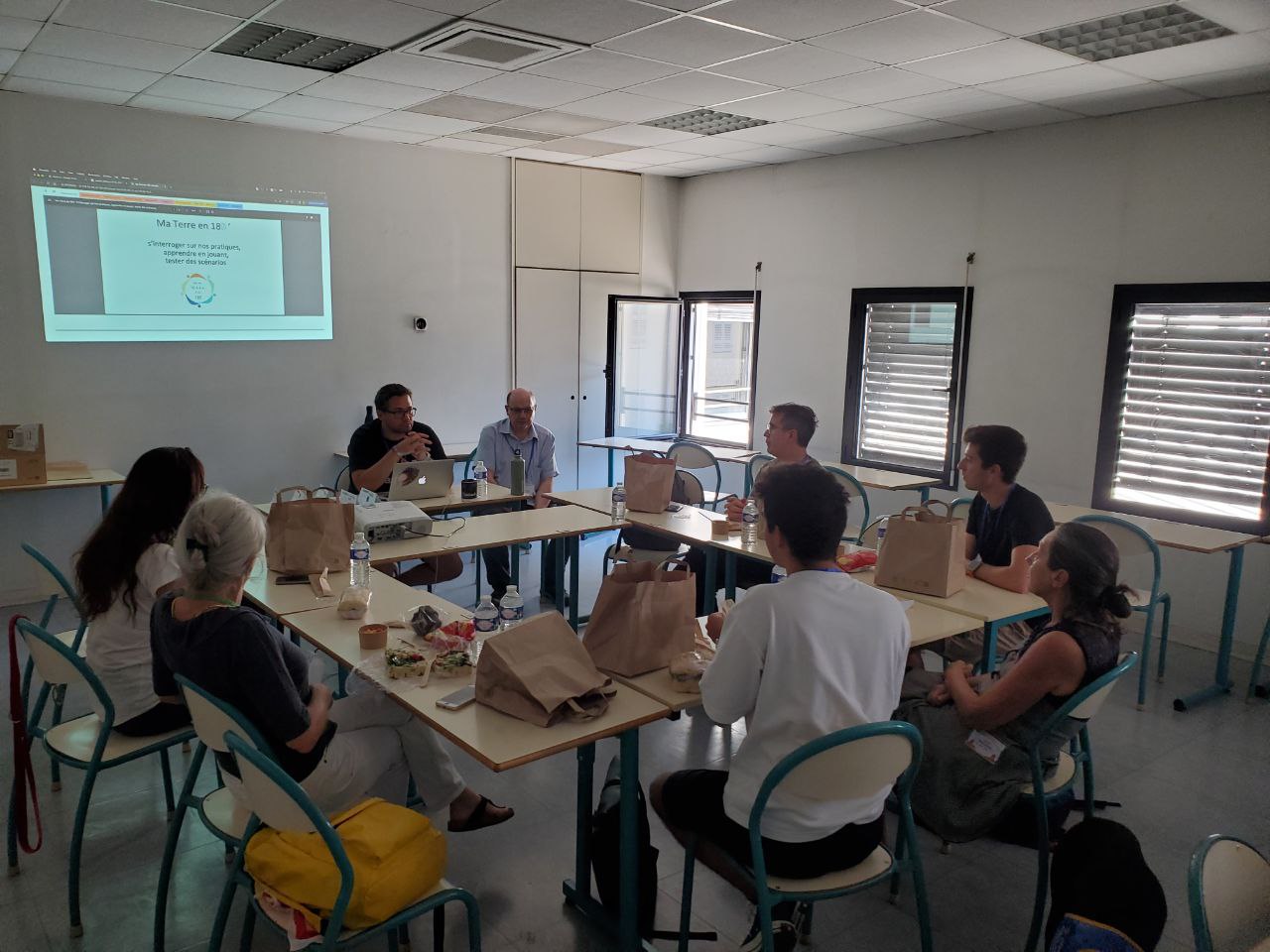}
  \caption{Pictures of three mini-workshops AST-01, AST-02 and AST-06
    among the seven that were organized. }
  \label{fig:pictures}
\end{figure}

\begin{figure}[t]
  \centering
  \includegraphics[width=0.8\textwidth]{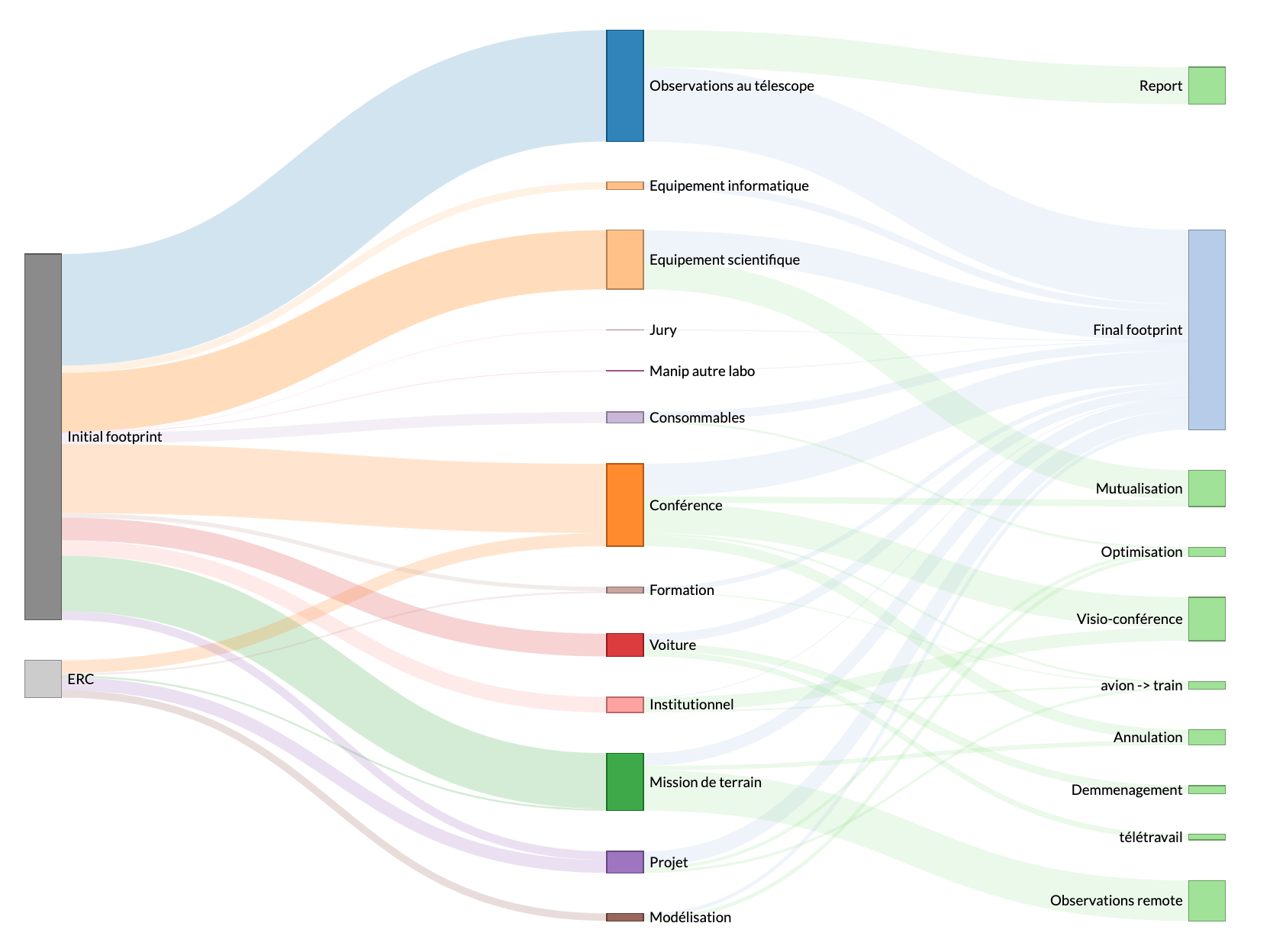}
  \caption{Decarbonization trajectory of the AST-06 group.}
  \label{fig:trajectory}
\end{figure}

During the SF2A session, seven teams for a total number of 45 participants
(Fig.~\ref{fig:pictures}) took part in the mini-workshops. The most effective
suggestions for reducing emissions included
(Fig.~\ref{fig:trajectory}):
\begin{list}{$-$}{\setlength{\itemsep}{0pt}\setlength{\parsep}{0pt}}{}
\item Systematically replacing flights with train travels for missions
  under 2000 km.
\item Promoting virtual meetings to avoid unnecessary travels.
\item Using archived data instead of organizing new carbon-intensive
  observations.
\item Extending the lifespan of scientific equipments.
\item Pooling long-distance travel for multiple purposes.
\end{list}

However, one of the main challenges highlighted was that achieving a
50\% reduction in emissions is extremely difficult without impacting
core research activities, such as telescope observations, which are
often deemed essential and difficult to cancel (Fig.~\ref{fig:results}).

\section{Perspectives and limitations}

\begin{figure}[t]
  \centering
  \includegraphics[width=0.8\textwidth]{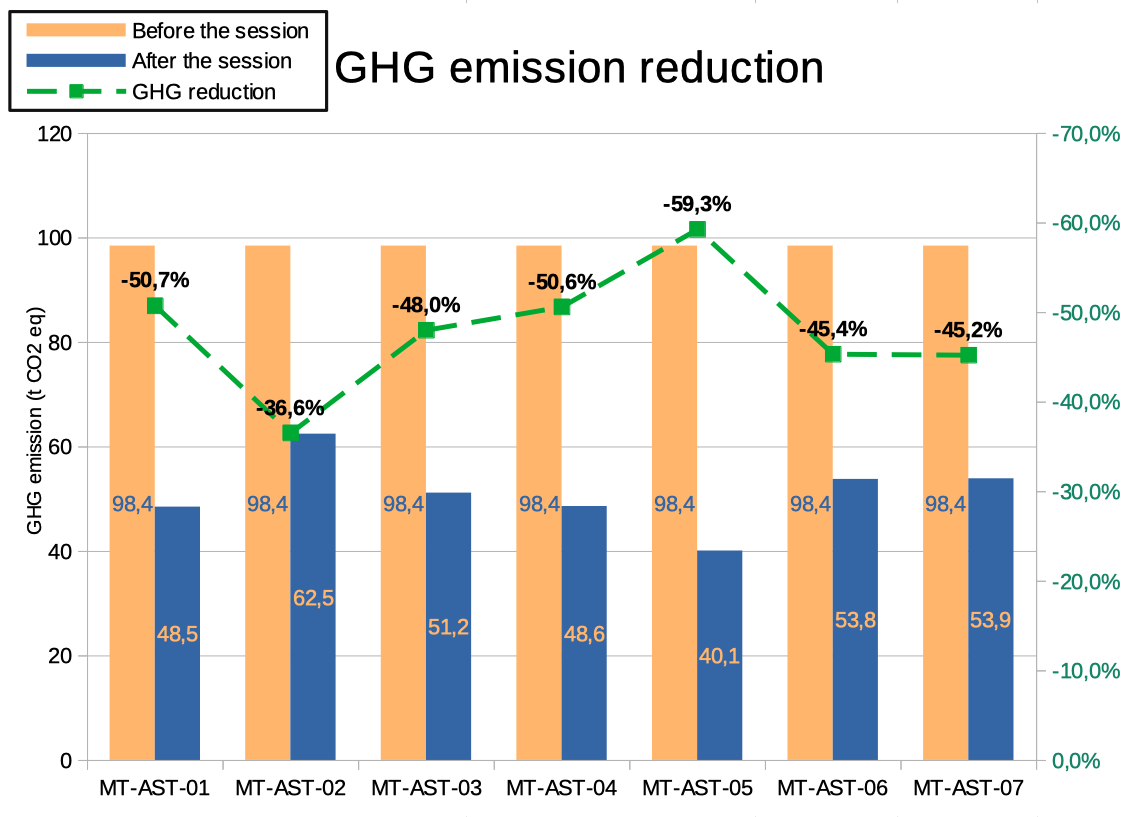}
  \caption{Summary of the trajectories of the various groups, which achieved emissions reductions ranging from 37\% to 59\%. }
  \label{fig:results}
\end{figure}

The results (Fig.~\ref{fig:results}) obtained during the SF2A session
were generally positive, although several limitations were
identified. For instance, some important emission-related topics were
not fully addressed like the role of reseearch infrastructure, and the
audience was already largely aware of the environmental issues, which
could bias the results.

Nevertheless, the project benefits from strong institutional support
in France, with partnerships from organizations such as CNRS, IRD,
INRAE, and Météo-France. Since 2021, more than 2000 participants have
taken part in over 300 workshops across France. To date, 25 virtual
teams are proposed and cover many scientific domains.

There are plans to expand the present serious game on a larger scale,
by testing the method on a broader, less familiar audience to assess
its real impact on environmental awareness, by adapting the serious
game to other scientific fields and cultural contexts, such as Vietnam
and the United States, by collaborating with specialists in cognitive,
social, and environmental sciences to refine and enhance the tool's
impact.

\section{Conclusion}

The "\emph{My Earth in 180 Minutes}'' serious game offers an innovative
approach to encourage low carbon activities in astronomy research
labs. It allows participants to model the impact of various strategies
on GHG emissions while raising awareness of the environmental
consequences of professional practices. Expanding this tool to other
fields and cultures promises new perspectives for further integrating
ecological transition into academic environments. In addition, this
serious game could be used as a programming tool to forecast emissions
for the coming years and that it canalso be well adapted to work on social justice.

For more information, visit the project’s official website: \url{https://materre.osug.fr}.

\begin{acknowledgements}
We would like to warmly thank the participants to the seven
mini-workshops, namely :
\begin{list}{$-$}{\setlength{\itemsep}{0pt}}
\item A.~Berdeu, A.~Jardin-Blicq, O.~Venot, N.~Brucy,
  Y.~Bernard, L.~d'Hendecourt, C.~Bergez
\item R.~Lenoble, V.~Hill, A.~Marchaudon, M.~Motte, M.~Tanious,
  R.~Lenoble, V.~Chevalier, R.~Peralta
\item V.~Durepaire, D.~Barret, S.~Lescaudron, L.~Michel-Dansac,
  I.~Boisse, A.~Schneeberger
\item M.~Zannese, G.~Chaverot, R.-M.~Ouazzani, C.~Dubois, F.~Antoine,
  J.-F.~Gonzalez, I.~Vaughlin
\item P.~Huet, L.~Cros, B.~Carry, A.~Borderies, A.~Niemiec,
  M.~Bethermin
\item C.~Gry, J.-P.~Berger, P.~Larue, S.~Brau Nogue, R.~Lallemand,
  L.~Tasca
\item M.~Mondelin, S.~Bontemps, M.~Van der Swaelmen, A.~Guseva,
  A.~Astoul, P.-A.~Desrotou
\end{list}

We would like to gratefully acknowledge the support of the CNRS, IRD, UGA, AMU, OSUG and LAM. We would also like to thank the support of the scientific organising committee and the local organising committee of the Journées Françaises de la SF2A, which enabled us to share this tool with our community.
\end{acknowledgements}

\end{document}